\documentclass[fleqn,usenatbib]{mnras}

\usepackage{newtxtext,newtxmath}

\usepackage[T1]{fontenc}
\usepackage{ae,aecompl}

\usepackage{graphicx}	
\usepackage{amsmath}	

\title[ASAS-SN Catalog of Variable Stars VII]{The ASAS-SN Catalog of Variable Stars VII: Contact Binaries are Different Above and Below the Kraft Break}

\author[T. Jayasinghe et al.]{T. Jayasinghe$^{1,2}$\thanks{E-mail: jayasinghearachchilage.1@osu.edu},
K. Z. Stanek$^{1,2}$,
C. S. Kochanek$^{1,2}$,
B. J. Shappee$^{3}$,
\newauthor 
M. H. Pinsonneault$^{1}$,
T. W. -S. Holoien$^{4}$,
Todd A. Thompson$^{1,2,5}$,
J. L. Prieto$^{6,7}$,
\newauthor
M. Pawlak$^{8}$,
O. Pejcha$^{8}$,
G. Pojmanski$^{9}$,
S. Otero$^{10}$,
N. Hurst$^{11}$,
D. Will$^{1,11}$
\\
$^{1}$Department of Astronomy, The Ohio State University, 140 West 18th Avenue, Columbus, OH 43210, USA\\
$^{2}$Center for Cosmology and Astroparticle Physics, The Ohio State University, 191 W. Woodruff Avenue, Columbus, OH 43210, USA\\
$^{3}$Institute for Astronomy, University of Hawaii, 2680 Woodlawn Drive, Honolulu, HI 96822,USA\\
$^{4}$Carnegie Observatories, 813 Santa Barbara Street, Pasadena, CA 91101, USA\\
$^{5}$Institute for Advanced Study, Princeton, NJ, 08540\\
$^{6}$N\'ucleo de Astronom\'ia de la Facultad de Ingenier\'ia y Ciencias, Universidad Diego Portales, Av. Ej\'ercito 441, Santiago, Chile\\
$^{7}$Millennium Institute of Astrophysics, Santiago, Chile\\
$^{8}$Institute of Theoretical Physics, Faculty of Mathematics and Physics, Charles University, Czech Republic\\
$^{9}$Warsaw University Observatory, Al Ujazdowskie 4, 00-478 Warsaw, Poland\\
$^{10}$The American Association of Variable Star Observers, 49 Bay State Road, Cambridge, MA 02138, USA\\
$^{11}$ASC Technology Services, 433 Mendenhall Laboratory 125 South Oval Mall Columbus OH, 43210, USA\\
}

\date{Accepted XXX. Received YYY; in original form ZZZ}

\pubyear{2019}

\begin{document}
\label{firstpage}
\pagerange{\pageref{firstpage}--\pageref{lastpage}}
\maketitle

\begin{abstract}
We characterize ${\sim} 71,200$ W UMa type (EW) contact binaries, including ${\sim} 
12,600$ new discoveries, using ASAS-SN $V$-band all-sky light curves along with archival data from 
Gaia, 2MASS, AllWISE, LAMOST, GALAH, RAVE, and APOGEE. There is a clean break in the EW 
period-luminosity relation at $\rm \log (\rm P/d){\simeq}-0.30$, separating the longer period 
early-type EW binaries from the shorter period, late-type systems. 
The two populations are even more cleanly separated in the space of period and effective 
temperature, by $\rm T_{eff}=6710\,K-1760\,K\,\log(P/0.5\,d)$. Early-type and late-type
EW binaries follow opposite trends in $\rm T_{eff}$ with orbital period. For longer 
periods, early-type EW binaries are cooler, while late-type systems are hotter. We 
derive period-luminosity relationships (PLRs) in the $W_{JK}$, $V$, Gaia DR2 $G$, $J$, $H$, $K_s$ and $W_1$ bands  for the late-type and early-type EW 
binaries separated both by period and effective temperature, and by period alone. The dichotomy of contact 
binaries is almost certainly related to the Kraft break and the related changes in 
envelope structure, winds and angular momentum loss.  

\end{abstract}

\begin{keywords}
stars:variables --stars:binaries:eclipsing -- catalogues --surveys 
\end{keywords}



\section{Introduction}
Contact binaries are close binary systems whose components fill their Roche Lobes. The most abundant are the W Ursae Majoris (W UMa) variables that are characterized by having nearly equal primary and secondary eclipse depths and orbital periods of ${\sim}0.2-1\, \rm d$. Given that both stars overflow their Roche lobes, the orbital periods of W UMa variables are closely related to the mean stellar densities \citep{1983ApJ...268..368E}. As a result, these contact binaries follow a period-luminosity relation, which can yield distances accurate to $10\%$ \citep{1994PASP..106..462R,2016ApJ...832..138C,2018ApJ...859..140C}. W UMa variables show little color variability and similar eclipse depths, so the component stars have similar effective temperatures and are in thermal contact \citep{2003ASPC..293...76W}. In most variable star catalogues, W UMa variables are assigned the GCVS/VSX \citep{2017ARep...61...80S,2006SASS...25...47W} classification of `EW'. 

W UMa variables are abundant in the Galaxy, and the advent of wide field surveys, such as the All-Sky Automated Survey (ASAS; \citealt{2002AcA....52..397P}), the Optical Gravitational Lensing Experiment (OGLE; \citealt{2003AcA....53..291U}), the Northern Sky Variability Survey (NSVS; \citealt{2004AJ....127.2436W}), MACHO \citep{1997ApJ...486..697A}, EROS \citep{2002A&A...389..149D}, the Catalina Real-Time Transient Survey (CRTS; \citealt{2014ApJS..213....9D}), the Asteroid Terrestrial-impact Last Alert System (ATLAS; \citealt{2018PASP..130f4505T,2018arXiv180402132H}), have yielded ${\gtrsim}10^5$ such variables. Using the ASAS catalog, \citet{2006MNRAS.368.1319R} estimated an abundance relative to FGK stars of $0.2\%$. Given their high occurrence rates, they can also be used to study Galactic structure \citep{1997AJ....113..407R}. 

Contact binaries play a significant role in stellar evolution and maybe the progenitors for objects such as blue stragglers \citep{2006ApJ...646.1160A,2009MNRAS.395.1822C} and Oe/Be stars \citep{2010NewAR..54...45E,2013ApJ...764..166D}. In one case, a contact binary was observed to evolve into a stellar merger \citep{2011A&A...528A.114T}. Thus, the study of the formation and evolution of contact binaries will improve our understanding of binary mergers and stellar evolution.

The detached-binary channel is considered to be crucial for the formation of contact binaries \citep{1986IAUS..118..159R,2007ApJ...662..596L,2014MNRAS.438..859J}. In this channel, a close detached binary evolves to Roche lobe overflow and then to contact either through the evolutionary expansion of the components \citep{1976ApJS...32..583W}  or through angular momentum loss by magnetic braking \citep{1982A&A...109...17V}. Studies of chromospherically active binaries have shown that they are losing angular momentum and evolving towards shorter orbital periods, making them good candidates for the progenitors of contact binaries \citep{2006MNRAS.373.1483E}. \citet{2006MNRAS.368.1311P} studied the eclipsing binaries in the ASAS catalog and noted a deficiency of close detached binaries with periods $\rm P<1$ d compared to the number needed to produce the observed number of contact binaries. It seems likely that many contact binaries form in triple systems, where the Kozai-Lidov mechanism \citep{1962P&SS....9..719L,1962AJ.....67..591K} drives the evolution towards becoming contact binaries \citep{2001ApJ...562.1012E}. The formation time-scale of contact binaries from a single starbust in the detached-binary channel has a large dynamic range (${\sim}1$ Myr to ${\sim}15$ Gyr, \citealt{2014MNRAS.438..859J}), which can explain the existence of very young contact binaries ($<10$ Myr) \citep{2011AJ....142...60V}.

W UMa contact binaries are observationally classified into A-type and W-type systems. The primary eclipse in the A-type systems corresponds to the transit of the secondary across the primary, whereas W-type systems are those whose secondaries are occulted by the primary. The sub-types also have different spectral types. A-type systems have A-F spectra and W-type systems have G-K spectra \citep{2003ASPC..293...76W}. Thus, the two sub-types appear to be separated in temperature, with A-type systems having temperatures ${\gtrsim}6000$ K \citep{1974AcA....24..119R}. The massive component of an A-type system is hotter than the less massive component and the opposite is true for W-type systems \citep{2013MNRAS.430.2029Y,1970VA.....12..217B}. The formation mechanism of A-type contact binaries in the pre-contact phase is dominated by the nuclear evolution of the more massive component ($1.8M_{\odot}<M<2.7M_{\odot}$) and the angular momentum evolution of the less massive component ($0.2M_{\odot}<M<1.5M_{\odot}$) \citep{2014MNRAS.437..185Y}. The pre-contact phases for the A-type systems typically end as the primary begins to evolve off the main-sequence. In the W-type systems, both components undergo efficient angular momentum loss and in most cases the angular momentum evolution is so rapid that the binary evolves into contact before the primary leaves the main-sequence \citep{2014MNRAS.437..185Y}. In this work, we will refer to the A-type systems as early-type EW binaries and the W-type systems as late-type EW binaries. The early/late types are related to the W/A sub-types but the classifications are not completely identical.

Main-sequence stars follow two distinct rotational regimes that are determined by how efficiently the stars lose angular momentum. Stars cooler than $\rm T_{eff}\lesssim6200\, K$ are slow rotators \citep{2013ApJ...776...67V}. These stars have thick convective envelopes, and rapidly lose their angular momentum due to magnetized winds. Hot stars with $\rm T_{eff}\gtrsim6700$ K rotate rapidly \citep{2007A&A...463..671R} because they do not have thick convective envelopes, and angular momentum loss through magnetized winds becomes very inefficient. The transition from the slowly-rotating main-sequence stars to the rapidly rotating main-sequence stars occurs at ${\sim}1.3 M_{\odot}$ (early F spectral types) and is known as the Kraft break \citep{1967ApJ...150..551K}. The differences in angular momentum loss above and below the Kraft break, presumably drive the evolutionary difference between the early and late-type binaries.

In a series of papers, \citet{2018MNRAS.477.3145J,2019MNRAS.485..961J,2019arXiv190710609J,2019arXiv191014187J,2019arXiv190100005J}, we have been systematically identifying and classifying variables using data from The All-Sky Automated Survey for SuperNovae (ASAS-SN, \citealt{2014ApJ...788...48S, 2017PASP..129j4502K}). We have thus far discovered  ${\sim}220,000$ new variables and homogeneously classified both the new and previously known variables in the sample \citep{2019MNRAS.486.1907J}. Here, we analyze an all-sky catalogue of 71242 W UMa (EW) contact binaries in the ASAS-SN V-band data. In Section $\S2$, we summarize the ASAS-SN catalogue of EW binaries and the cross-matching to external photometric and spectroscopic catalogues. We analyze the sample of EW binaries with spectroscopic cross-matches and compare early-type and late-type systems in Section $\S3$. We derive period-luminosity relationships for these two sub-types in Section $\S 4$. The V-band light curves and other variability and photometric information for all of the ${\sim}71,200$ sources studied in this work are available online at the ASAS-SN variable stars database (\url{https://asas-sn.osu.edu/variables}). 

\section{The ASAS-SN Catalog of contact binaries}

In this work, we selected 71242 W UMa-type contact binary stars (EW) identified during our systematic search for variables, including new EW binaries in the Northern hemisphere and regions of the southern Galactic plane that were missed in the previous survey papers (Jayasinghe et al. 2020, in prep). Out of the 71242 EW binaries in this catalogue, 12584 (${\sim}18\%$) are new ASAS-SN discoveries. The ASAS-SN V-band observations used in this work were made by the ``Brutus" (Haleakala, Hawaii) and ``Cassius" (CTIO, Chile) quadruple telescopes between 2013 and 2018. Each ASAS-SN V-band field is observed to a depth of $V\lesssim17$ mag. The field of view of an ASAS-SN camera is 4.5 deg$^2$, the pixel scale is 8\farcs0 and the FWHM is typically ${\sim}2$ pixels. ASAS-SN saturates at ${\sim} 10-11$ mag, but we attempt to correct the light curves of saturated sources for bleed trails (see \citealt{2017PASP..129j4502K}). The V-band light curves were extracted as described in \citet{2018MNRAS.477.3145J} using image subtraction \citep{1998ApJ...503..325A,2000A&AS..144..363A} and aperture photometry on the subtracted images with a 2 pixel radius aperture.  The APASS catalog \citep{2015AAS...22533616H} and the ATLAS All-Sky Stellar Reference Catalog \citep{2018ApJ...867..105T} were used for calibration. We corrected the zero point offsets between the different cameras as described in \citet{2018MNRAS.477.3145J}. The photometric errors were recalculated as described in \citet{2019MNRAS.485..961J}. 

Variable sources were identified and subsequently classified using two independent random forest classifiers plus a series of quality checks as described in \citet{2019MNRAS.486.1907J,2019arXiv190710609J}. We used the \verb"astropy" implementation of the Generalized Lomb-Scargle (GLS, \citealt{2009A&A...496..577Z,1982ApJ...263..835S}) periodogram and the \verb"astrobase" implementation \citep{astrob} of the Box Least Squares (BLS, \citealt{2002A&A...391..369K}) periodogram, which improves the completeness for eclipsing binaries, to search for periodicity over the range $0.05\leq P \leq1000$ days. We classified the eclipsing binaries into the GCVS/VSX photometric (sub-)classes: EW, EB and EA. EW (W UMa) binaries have light curves with minima of similar depths whereas EB ($\beta$-Lyrae) binaries tend to have minima of significantly different depths. The ratio of eclipse depths ($D_s/D_p)$ for most contact binaries are $D_s/D_p>0.8$, whereas most semi-detached systems have eclipses of different depths with $D_s/D_p<0.8$ \citep{2006MNRAS.368.1311P,2019MNRAS.486.1907J}. Most contact binaries are in thermal contact, but \citet{2006MNRAS.368.1311P} also noted systems with unequal minima, implying that some contact binaries are not in thermal contact, as was predicted by models of thermal relaxation oscillations (see, for e.g., \citealt{1976ApJ...205..208L,1976ApJ...205..217F,2005ApJ...629.1055Y}). Both the EW (contact) and EB (contact/semi-detached) binaries transition smoothly from the eclipse to the out-of-eclipse state. EA (Algol) binaries are detached systems where the exact onset and end of the eclipses are easily defined. These detached systems may or may not have a secondary minimum.

We cross-matched the EW binaries with Gaia DR2 \citep{2018arXiv180409365G} using a matching radius of 5\farcs0. The sources were assigned distance estimates from the Gaia DR2 probabilistic distance estimates \citep{2018AJ....156...58B} by cross-matching based on the Gaia DR2 \verb"source_id". A large majority of these sources (${\sim}86.8\%$) had distance estimates from Gaia DR2. We also cross-matched these sources to the 2MASS \citep{2006AJ....131.1163S} and AllWISE \citep{2013yCat.2328....0C,2010AJ....140.1868W} catalogues using a matching radius of 10\farcs0. We used \verb"TOPCAT" \citep{2005ASPC..347...29T} for this process. Following the cross-matching process, we calculated the absolute, reddening-free Wesenheit magnitude \citep{1982ApJ...253..575M,2018arXiv180803659L} for each source as 
\begin{equation}
    W_{JK}=M_{\rm K_s}-0.686(J-K_s) \,.
	\label{eq:wk}
\end{equation} For each source, we also calculate the total line of sight Galactic reddening $E(B-V)$ from the recalibrated `SFD' dust maps \citep{2011ApJ...737..103S,1998ApJ...500..525S}.

We cross-matched our sample with the APOGEE DR15 catalog \citep{2015AJ....150..148H, 2017AJ....154...94M}, the RAVE-on catalog \citep{2017ApJ...840...59C}, the LAMOST DR5 v4 catalog \citep{2012RAA....12.1197C} and the GALAH DR2 catalog \citep{2015MNRAS.449.2604D,2018MNRAS.478.4513B} using a matching radius of 5\farcs0. We identified 7169 matches to the EW binaries from the LAMOST ($94.0\%$), GALAH ($3.8\%$), RAVE ($2.1\%$) and APOGEE ($0.1\%$) spectroscopic surveys. 

\begin{figure*}
	\includegraphics[width=\textwidth]{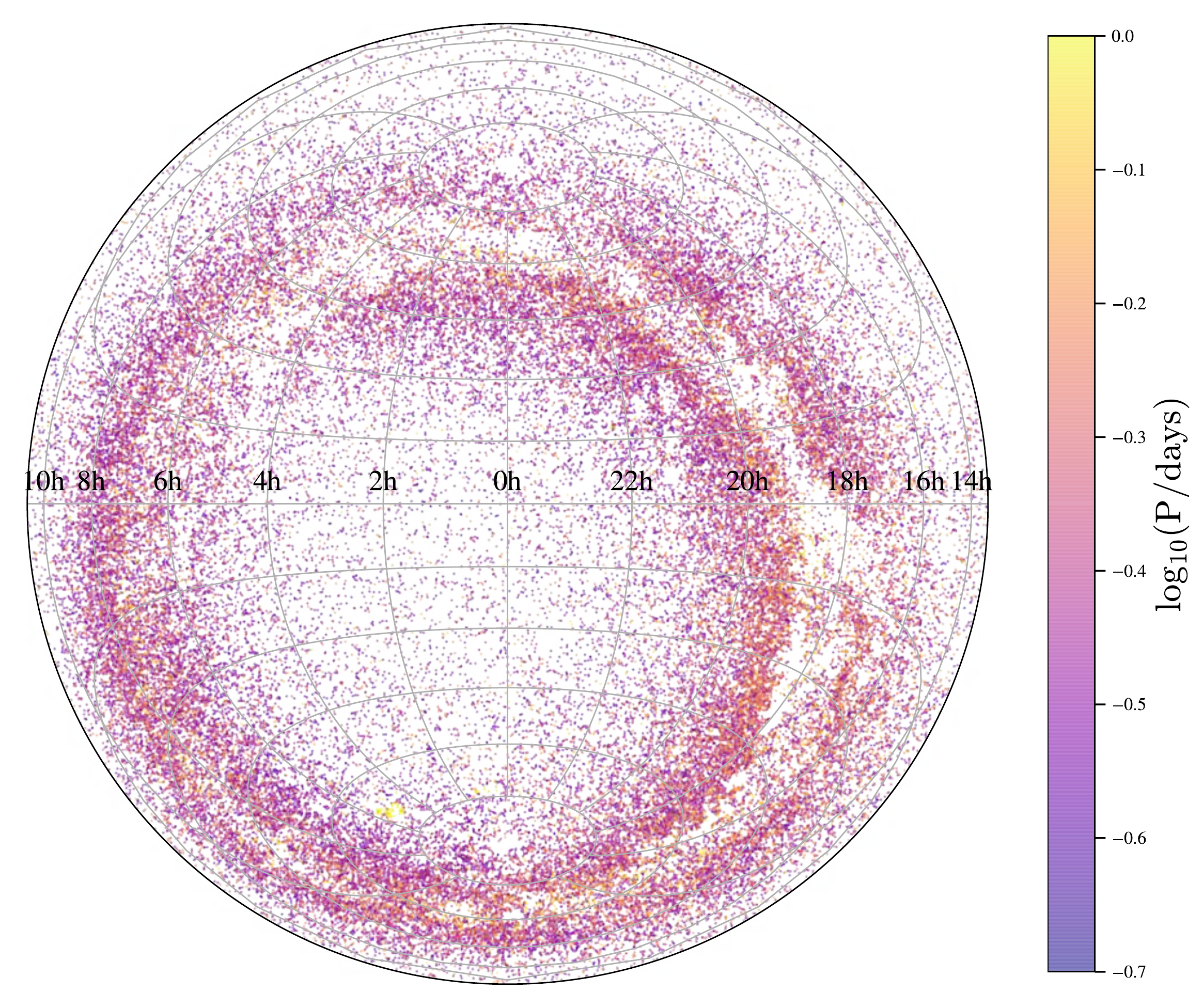}
    \caption{Projected distribution of the $\sim 71,200$ EW binaries in Equatorial coordinates (Lambert projection). The points are colored by period.}
    \label{fig:fig1}
\end{figure*}

The median V-band magnitude of the EW binary sample is $V{\sim}14.7$ mag. Classification probabilities of $\rm Prob>0.9$ are considered very reliable and ${\sim}93.4\%$ of our sample of contact binaries have $\rm Prob>0.9$. There are 21837 sources within $1\, \rm kpc$ but a considerable fraction (${\sim}69\%$) of the sources with Gaia DR2 distances are located further away. A large fraction have useful parallaxes, as ${\sim}69\%$ (${\sim}58\%$) of the sources have \verb"parallax"/\verb"parallax_error" $>5$ ($>10$). The median line-of-sight extinction to the EW binaries is $A_V {\sim}0.36$ mag, assuming $R_V=3.1$ dust \citep{1989ApJ...345..245C}. The sky distribution of the EW binaries in ASAS-SN, colored by their period, is shown in Figure \ref{fig:fig1}. 

\begin{figure*}
	\includegraphics[width=\textwidth]{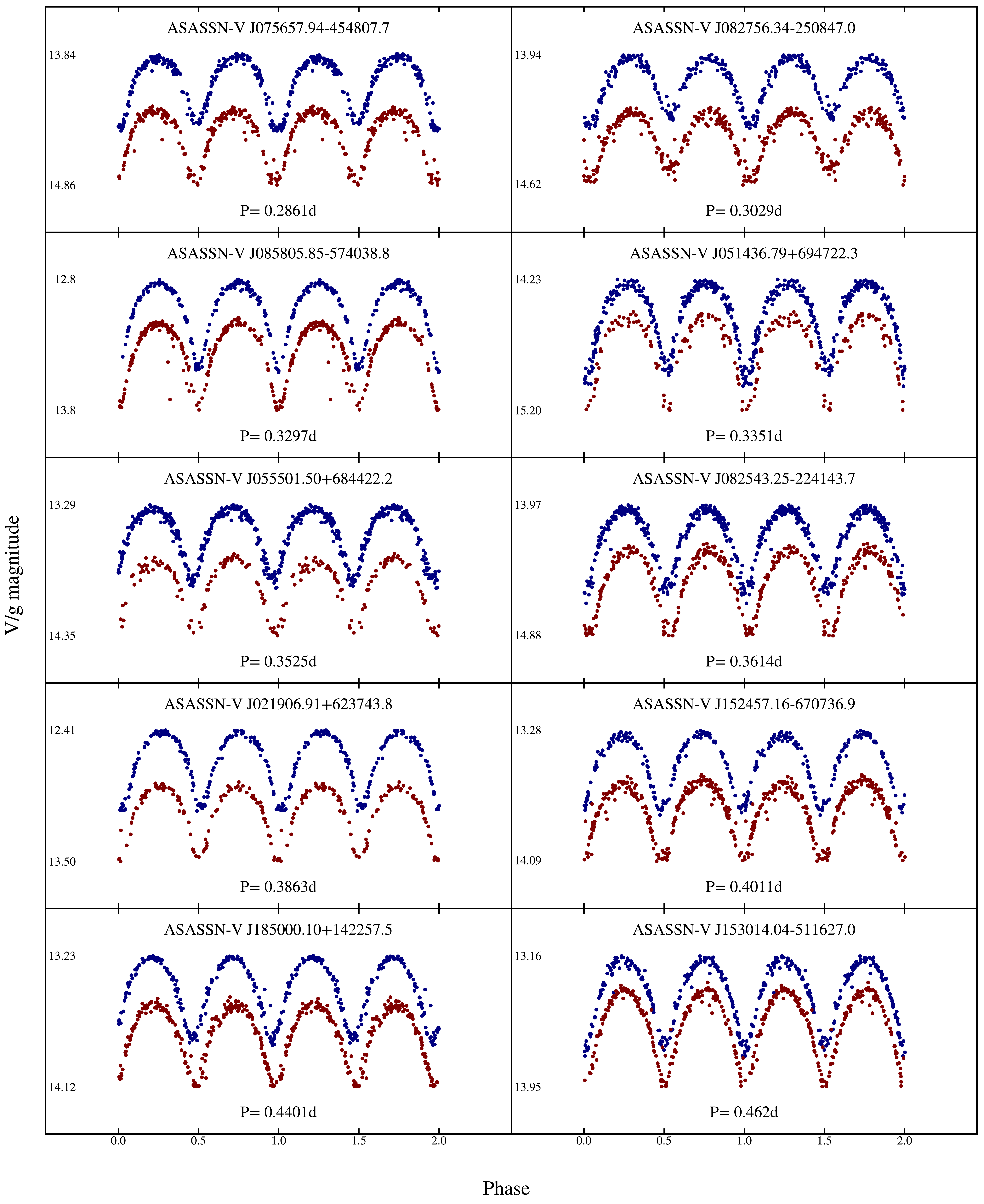}
    \caption{Phased ASAS-SN light curves for 10 late-type EW binaries. The light curves are scaled by their minimum and maximum V/g-band magnitudes. The blue (red) points are for the g(V) band data.}
    \label{fig:fig2}
\end{figure*}

\begin{figure*}
	\includegraphics[width=\textwidth]{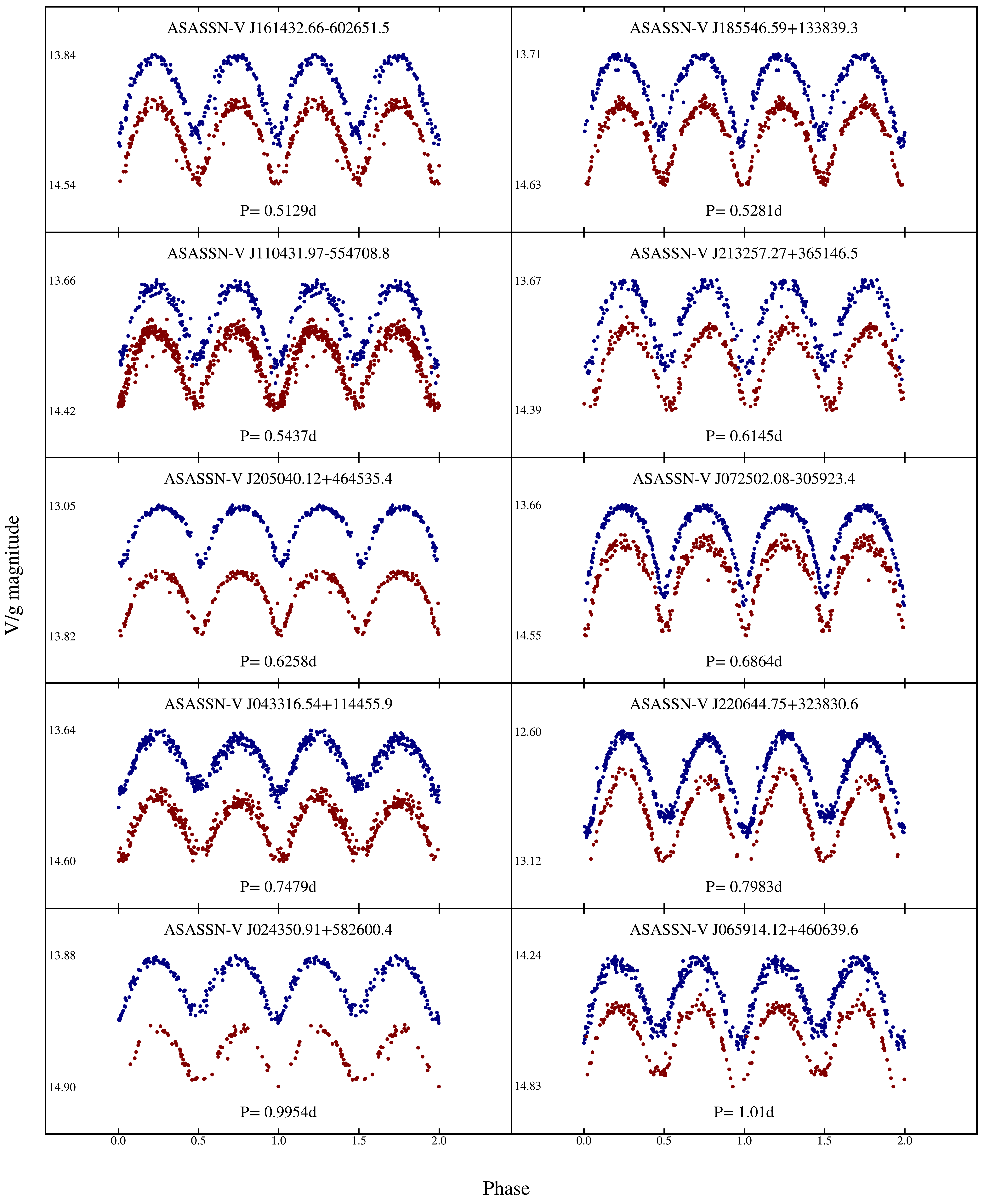}
    \caption{Phased ASAS-SN light curves for 10 early-type EW binaries. The format is the same as Figure \ref{fig:fig2}. }
    \label{fig:fig3}
\end{figure*}

We present the ASAS-SN light curves for 10 late-type and 10 early-type contact binaries to illustrate the light curve morphologies of these systems. The phased ASAS-SN V- and g-band light curves are shown in Figure \ref{fig:fig2} (late-type EW) and Figure \ref{fig:fig3} (early-type EW). The ratio of eclipse depths is similar for both the late-type and early-type binaries. We do not see substantial differences in the morphologies of the light curves between early-type and late-type contact binaries. The variations in the depths of the minima are generally only a few percent \citep{2003ASPC..293...76W}, making it challenging to distinguish the early and late-type systems using the ASAS-SN light curves.

The period distribution of the EW binaries is shown in Figure \ref{fig:fig4}. The EW binaries have a bimodal orbital period distribution, with the late-type systems (canonically defined as those with $\rm \log (\rm P/d) <-0.25$) having a median period of $\rm \log (\rm P/d) {\sim}-0.4$, and the early-type systems having a median period of $\rm \log (\rm P/d) {\sim}-0.15$. Most sources with periods $\rm P>1 d $ are likely $\beta$-Lyrae eclipsing binaries (EB), with nearly equal minima that are misclassified as contact (EW) binaries. We will discuss the differences between early-type and late-type contact binaries in Section $\S 3$ and derive period-luminosity relationships for these two sub-types in Section $\S 4$. 
\newpage 
\section{Early-type vs. Late-type W UMa contact binaries}
In previous studies, early-type and late-type EW binaries are usually separated on the basis of their period, with the early-type systems defined to have orbital periods $\log (\rm P/d) >-0.25$ (e.g.,\citealt{2018ApJ...859..140C}). It is also known that the early-type systems are fewer in number than the late-type systems \citep{2016MNRAS.457.4323P}. Figure \ref{fig:fig5} shows the Wesenheit $W_{JK}$ period-luminosity relationship (PLR) diagram for the EW binaries with classification probabilities of $\rm Prob>0.9$, $A_V<1$ mag and \verb"parallax"/\verb"parallax_error" $>10$.  In the sample of ASAS-SN contact binaries, we immediately see that the slope of the late-type PLR is steeper than that of the early type systems. \citet{2016MNRAS.457.4323P} studied a sample of early-type contact binaries in the Large Magellanic Cloud discovered by the OGLE survey \citep{2011AcA....61..103G,2016AcA....66..421P} and noted that the PLR for contact binaries are best described by two separate relations for the late-type and early-type systems. He found a shallower slope for the early-type contact binaries when compared to the late-type PLR. The PLRs for late-type EW binaries have been extensively studied \citep{1994PASP..106..462R,2016ApJ...832..138C,2018ApJ...859..140C}.

While the traditional period for separating early and late type EW systems is at $\rm \log (\rm P/d) =-0.25$, the clear minimum in the period distribution (Figure \ref{fig:fig4}) suggests that $\rm \log (\rm P/d) =-0.30$ is a better choice. In practice (see below), the two classes have some period overlap with early type systems having periods as short as $\rm \log (\rm P/d) =-0.40$ and late-type systems have periods as long as $\rm \log (\rm P/d) =-0.25$. Figure \ref{fig:fig5} shows the distribution of the systems in period and $W_{JK}$ and, we see a clear break in the slope at a period of $\rm \log (\rm P/d) {\sim}-0.30$. We will fit models for the PLR in Section $\S 4$ after improving the separation between the early-type and late-type systems. 

\citet{2014MNRAS.438..859J} found that the upper limit of the initial orbital period for binaries that come into contact (${\sim}3-4.2$ d) is significantly longer than the upper limit of the observed EW period distribution (${\sim}1-2$ d). Thus, the orbital period distribution for the EW binaries reflects significant orbital shrinkage compared to a zero-age binary population.  This implies that the physics of the merger process must be responsible for shaping the observed PLRs of the early-type and late-type systems. A successful theory has to consider the evolution from a detached system to a contact system. 

\begin{figure}
	\includegraphics[width=0.5\textwidth]{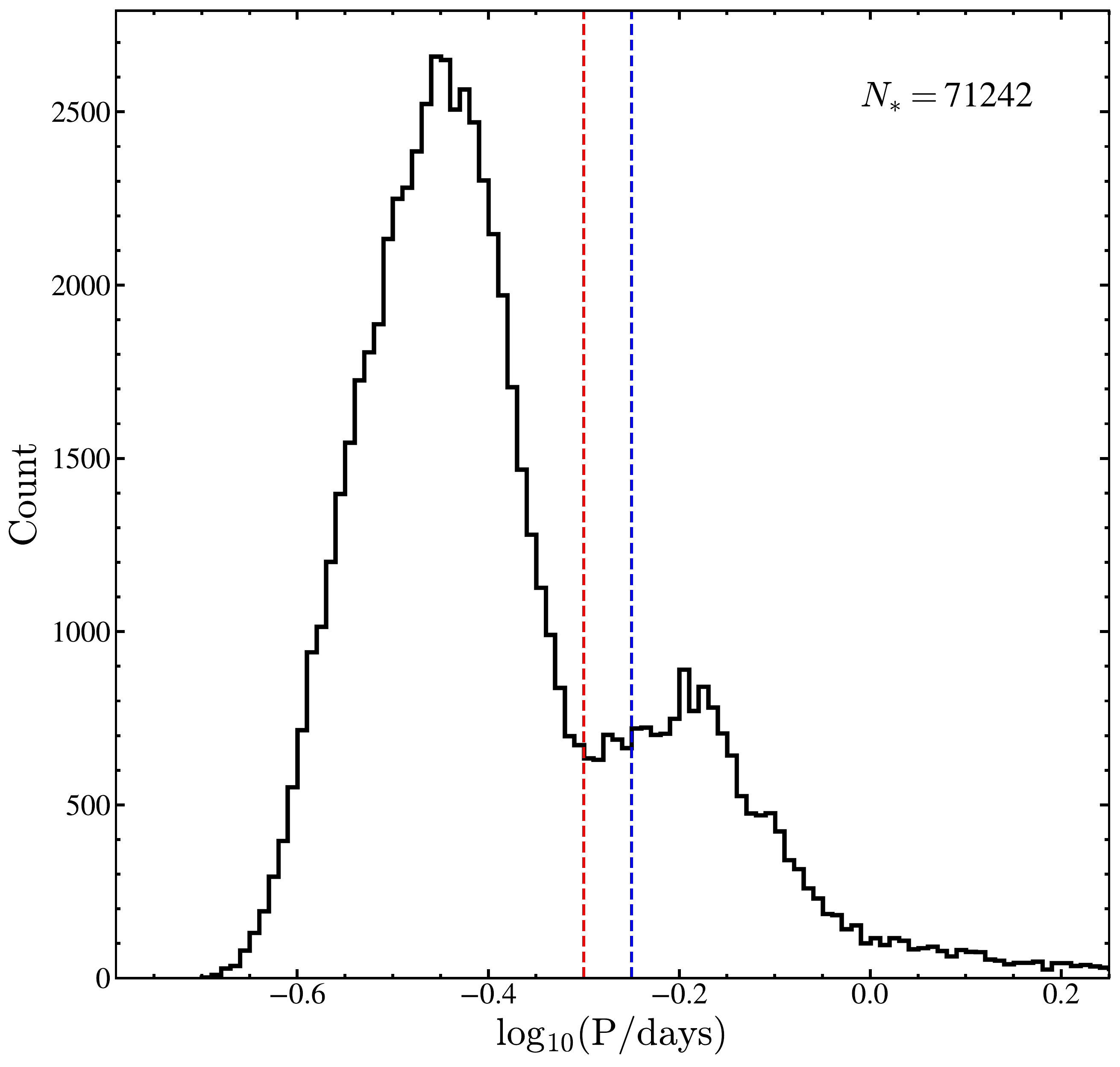}
    \caption{The distribution of orbital periods for the $\sim 71,200$ EW binaries. The usual period cut separating early-type and late-type systems of $\rm \log (\rm P/d) =-0.25$ is shown as a dashed blue line. A revised period cut of $\rm \log (\rm P/d) =-0.30$ based on the period distribution of the EW binaries in ASAS-SN is shown as a dashed red line.}
    \label{fig:fig4}
\end{figure}

\begin{figure}
	\includegraphics[width=0.5\textwidth]{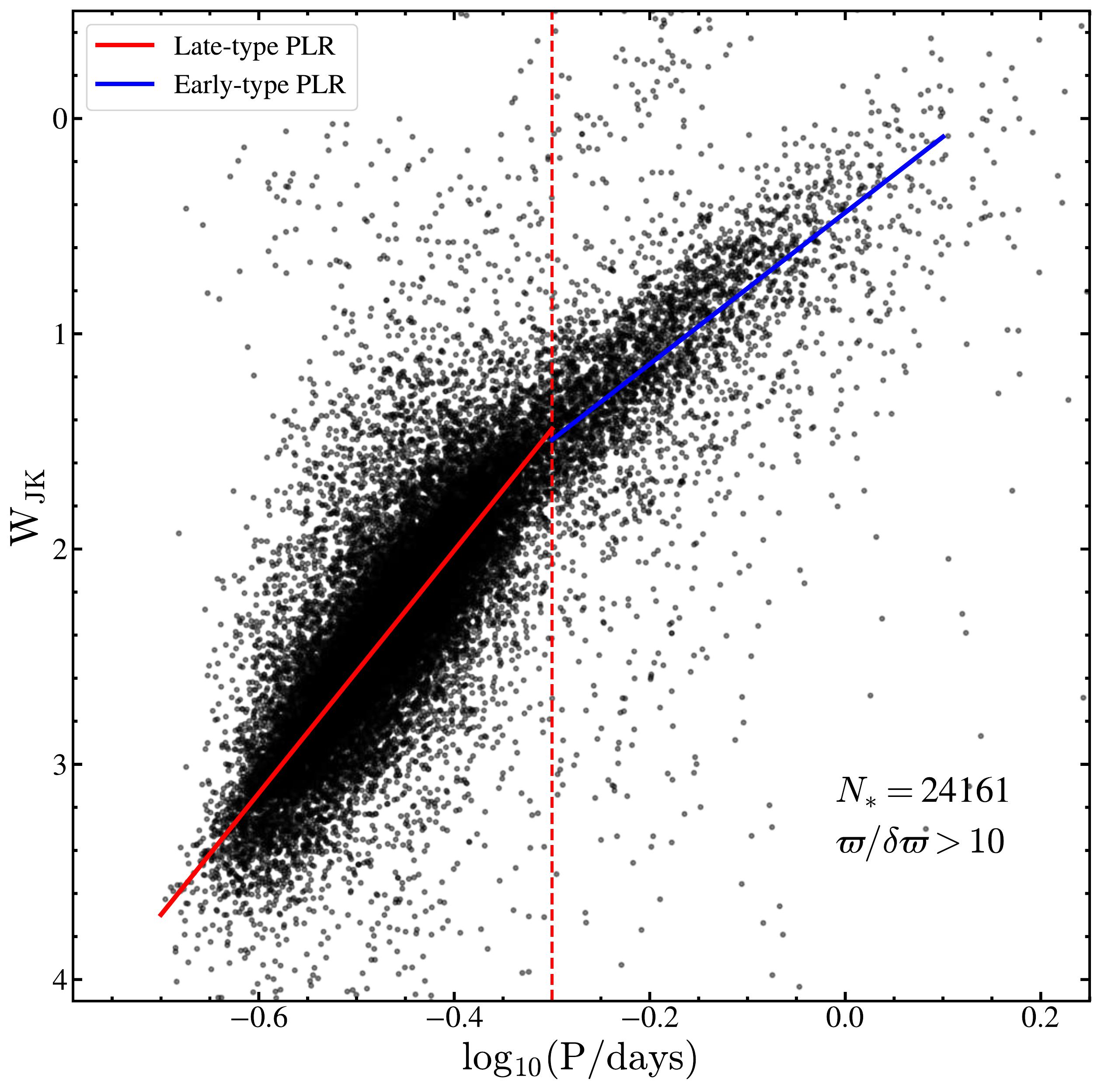}
    \caption{The Wesenheit $W_{JK}$ PLR diagram for the EW stars with $\rm Prob>0.90$, $A_V<1$ mag and parallaxes better than $10\%$. The fitted PLRs for the late-type and early-type contact binaries are shown as red and blue lines respectively. }
    \label{fig:fig5}
\end{figure}

\begin{figure*}
	\includegraphics[width=\textwidth]{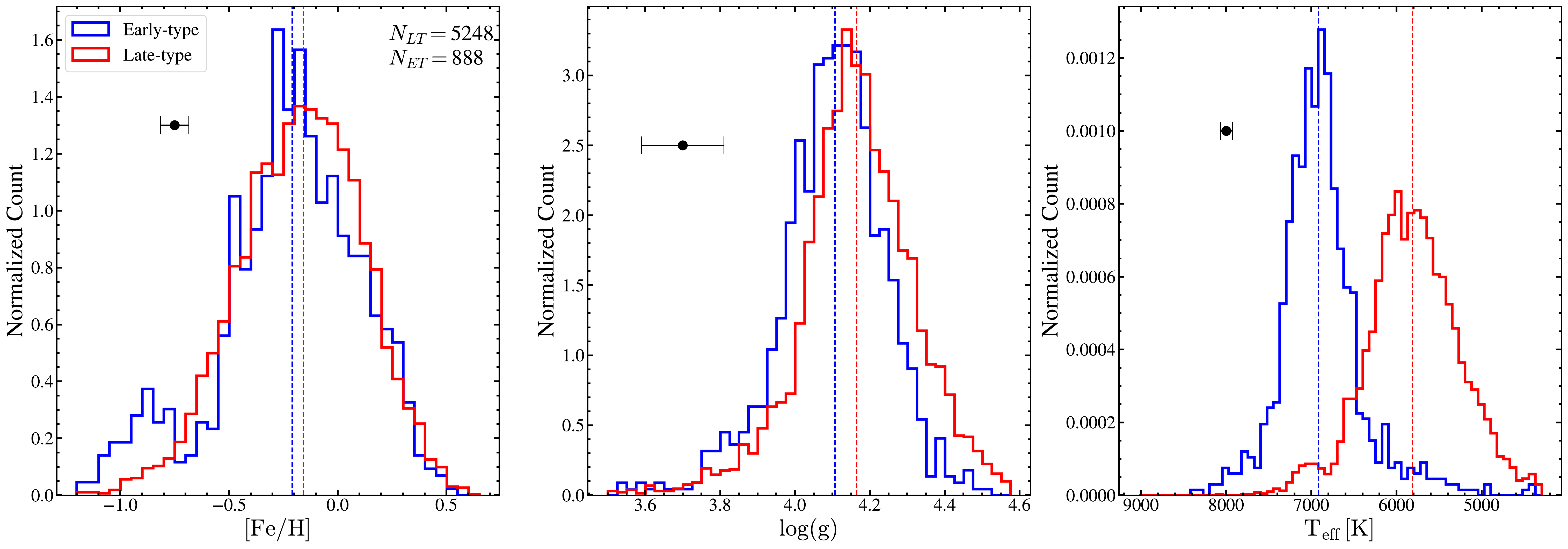}
    \caption{Distributions of the late-type (red) and early-type (blue) EW contact binaries with $\rm Prob>0.90$ in $\rm [Fe/H]$, $\rm \log (g)$, and $\rm T_{eff}$. The median value for each parameter is illustrated with a dashed line. The average uncertainty for each parameter is shown in black. The excess of early-type binaries with $\rm [Fe/H]<-0.8$ is due to misclassified RRc variables. }
    \label{fig:fig6}
\end{figure*}

Figure \ref{fig:fig6} shows the distributions in effective temperature $\rm T_{eff}$, surface gravity $\rm \log (g)$ and metallicity $\rm [Fe/H]$ for the early-type and late-type contact binaries with $\rm Prob>0.90$ in the APOGEE, LAMOST, GALAH and RAVE surveys. The distributions of the late-type ($N=5248$) and early-type ($N=888$) contact binaries in $\rm \log (g)$ and $\rm [Fe/H]$ are similar. There appears to be an excess of early-type binaries with $\rm [Fe/H]<-0.8$. However, upon further inspection, we find that a substantial fraction of these low-metallicity sources are actually misclassified overtone RR Lyrae (RRc) variables that were assigned twice their true pulsational period in the ASAS-SN pipeline. RRc variables are sometimes confused with EW variables because of their symmetric light curve morphologies. We discard early-type binaries with $\rm [Fe/H]<-0.8$ to minimize contamination.

\begin{figure*}
	\includegraphics[width=\textwidth]{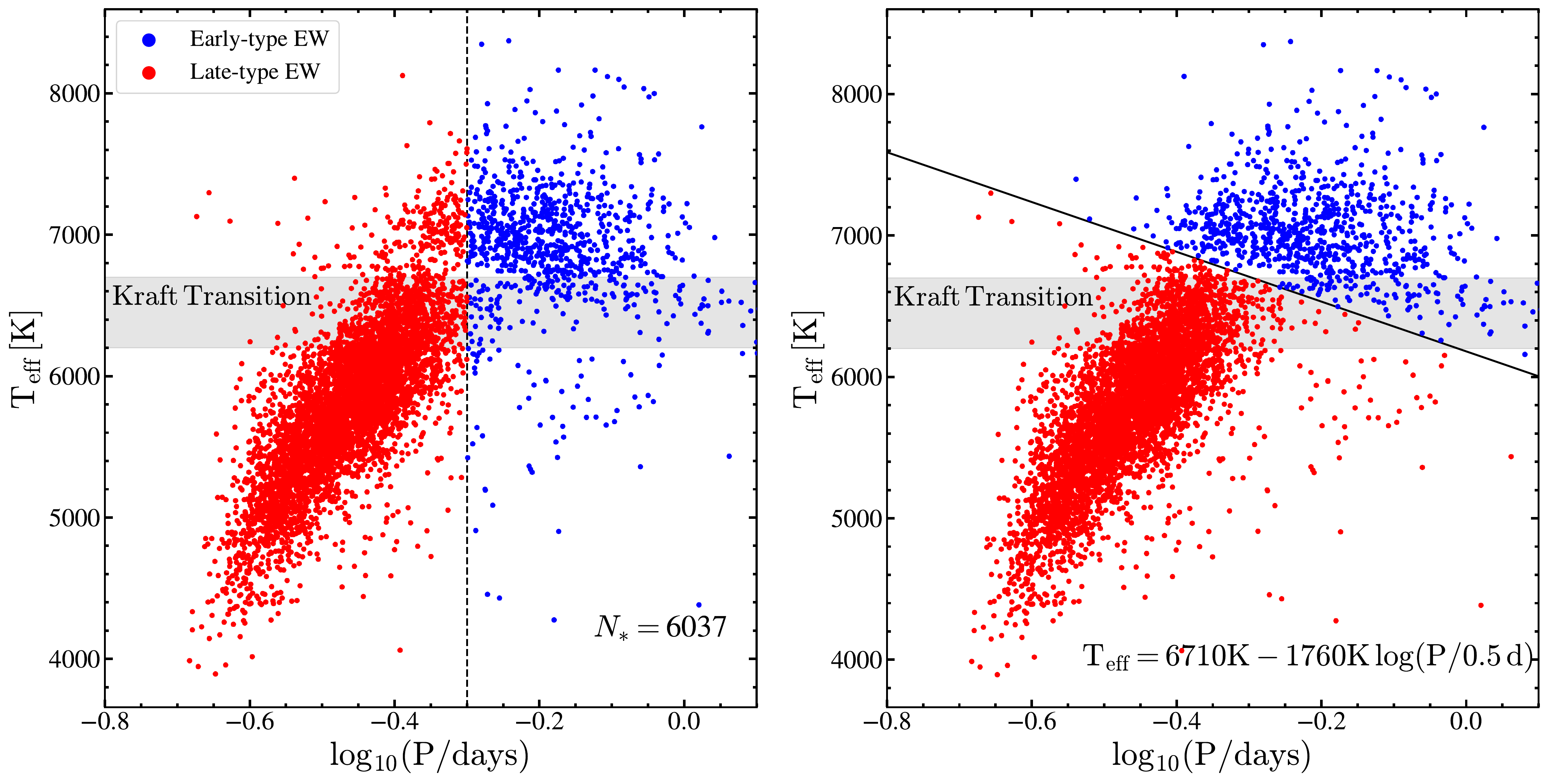}
    \caption{$\rm T_{eff}$ vs. $\rm \log (\rm P/d)$ for late-type (red) and early-type (blue) EW binaries separated by period (left) and a discriminant in $\rm T_{eff}$- $\rm \log (\rm P/d)$ (right). The Kraft transition from slow to fast rotation occurs over the temperature range $6200-6700\, \rm K$ and is shaded in gray. }
    \label{fig:fig7}
\end{figure*}

The temperature distributions are, of course, quite different, but the separation of the two populations is very striking in the space of $\rm \log (\rm P/d)$ and $\rm T_{eff}$, as shown in Figure \ref{fig:fig7}. The increase in temperature with period is well known for the late-type systems, going back to the observation that longer period systems are bluer. The trend reverses for the late-type systems, which has not previously been observed. \citet{2017RAA....17...87Q} noted a trend but dismissed it as scatter. Still more striking is that there is a clean break between the two populations which we can empirically model as \begin{equation}
    \rm T_{eff}=6710K-1760K\,\log \rm (P/0.5\,d) \,.
	\label{eq:tefflogp}
\end{equation} The traditional cut in period is an approximation to the actual separation of the two populations but a period separation of $\rm \log (\rm P/d) =-0.30$ at the minimum of the period distribution is a better ``average'' choice than the traditional $\rm \log (\rm P/d) =-0.25$. We use Equation \ref{eq:tefflogp} to separate the two populations in the spectroscopic sample. 

The existence of the Kraft break \citep{1967ApJ...150..551K} implies substantial changes in the envelope structure, winds and angular momentum loss for stars on the main-sequence. Stars above the Kraft break are hotter and rotate more rapidly than those below the Kraft break. The transition from slow to fast rotation occurs over the temperature range $6200-6700\, \rm K$ and it cannot be a coincidence that the split between early and late type contact binaries occurs at a similar temperature. Formation models for these systems generally invoke changes in the efficiency of angular momentum loss on the main sequence \citep{2014MNRAS.437..185Y} which is exactly the physics leading to the Kraft break. The remarkable feature of Figure \ref{fig:fig7} is the existence of a clear gap between early and late type systems, which seems not to be predicted in any models.

\begin{figure*}
	\includegraphics[width=\textwidth]{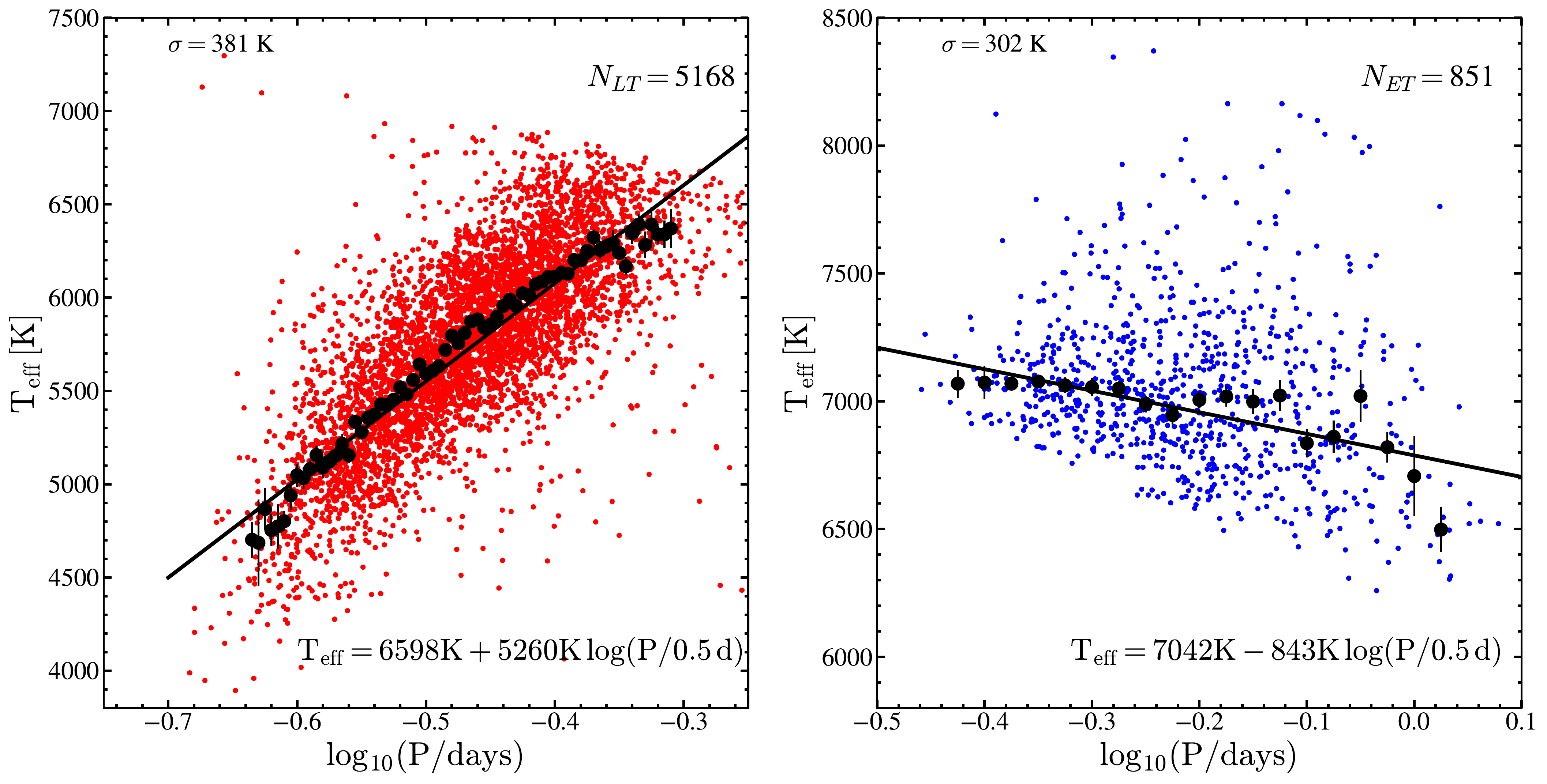}
    \caption{$\rm T_{eff}$ vs. $\rm \log (\rm P/d)$ for the late-type (left) and early-type (right) EW binaries. Linear fits to the binned data are shown in black.}
    \label{fig:fig8}
\end{figure*}

\begin{figure*}
	\includegraphics[width=\textwidth]{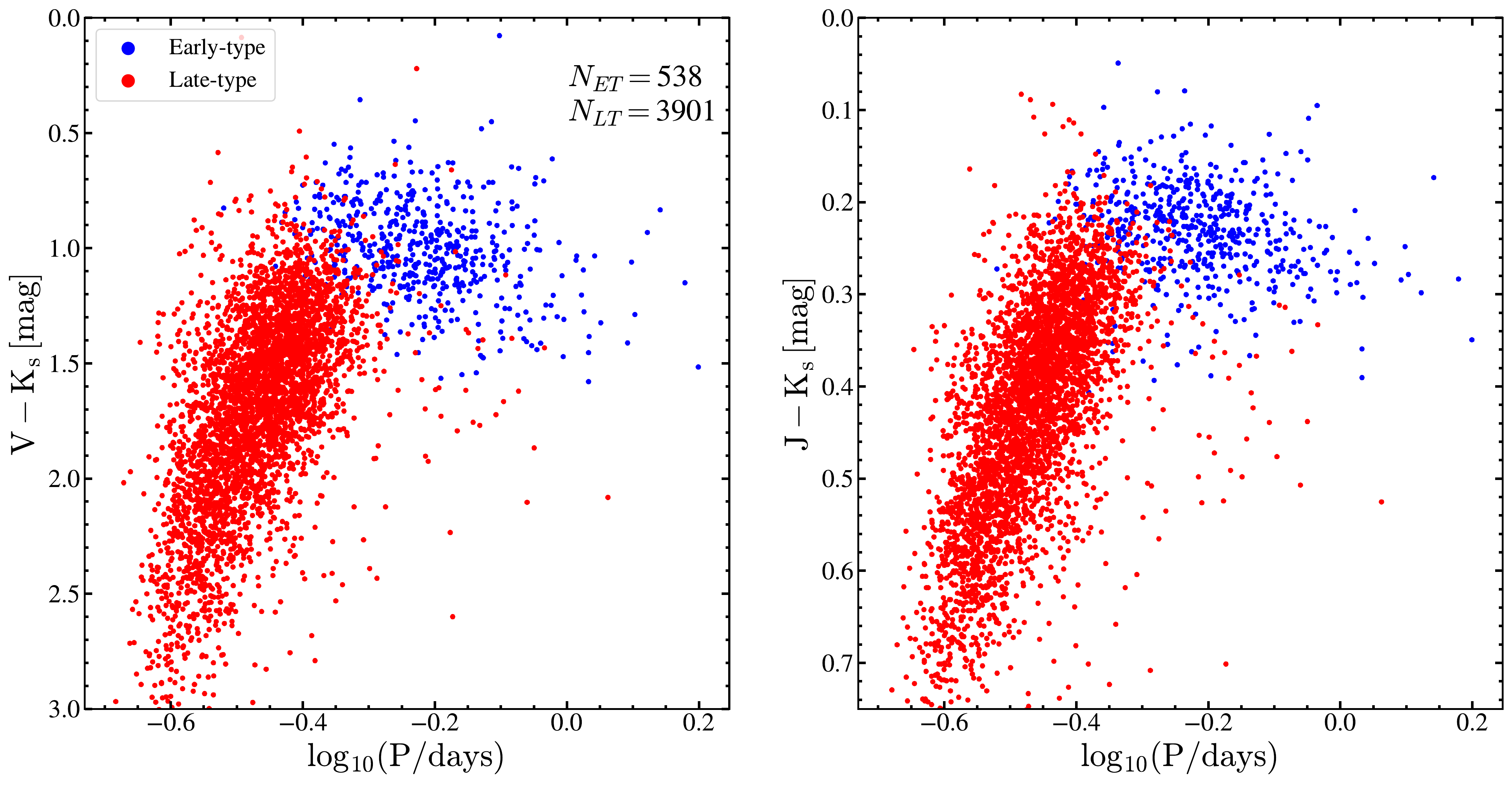}
    \caption{$V-K_s$ (left) and $J-K_s$ (right) vs. $\rm \log (\rm P/d)$ for the spectroscopically separated late-type (red) and early-type (blue) EW binaries separated in period-temperature space using equation \ref{eq:tefflogp}, with $\rm Prob>0.9$ and $A_V<0.5$ mag.}
    \label{fig:fig9}
\end{figure*}

If we separate the systems using Equation \ref{eq:tefflogp}, we find period-temperature relations of \begin{equation}
    \rm T_{eff} (LT)=6598(\pm23)K+\rm 5260(\pm116)K\,\log_{10}(P/0.5\,d),
	\label{eq:fehfund}
\end{equation} for the late-type systems and\begin{equation}
    \rm T_{eff} (ET)=7041(\pm28)K\rm -843(\pm164)K\,\log_{10}(P/0.5\,d),
	\label{eq:fehot} 
\end{equation} for the early-type systems. These relationships both have large scatter ($\sigma{\gtrsim}300 \, \rm K$), but it is clear that the slopes are not only very different but reverse in sign (Figure \ref{fig:fig8}). This can also be seen in the period-color distributions shown in Figure \ref{fig:fig9}. Historically, the period-color relation of late-type systems have been well characterized \citep{1961RGOB...31..101E,1967MmRAS..70..111E}. Here we see that the early-type systems become redder (cooler) with increasing orbital period, which is the reverse of the well-known correlations for the late-type systems.

\begin{figure*}
	\includegraphics[width=\textwidth]{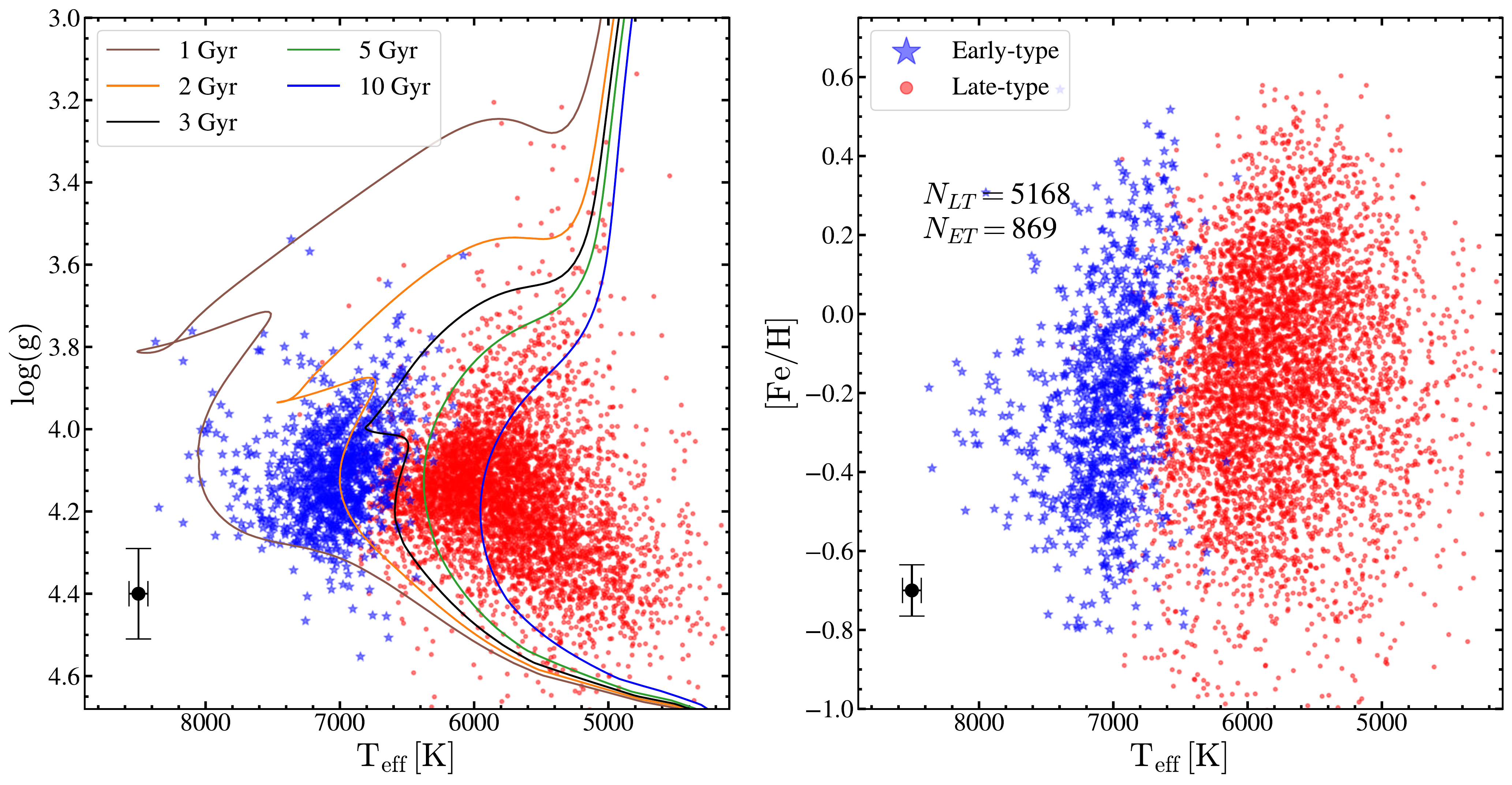}
    \caption{Distributions of the early-type and late-type contact binaries in $\rm \log (g)$ vs. $\rm T_{eff}$ (left) and $\rm T_{eff}$ vs. $\rm [Fe/H]$ (right). MIST isochrones \citep{2016ApJ...823..102C,2016ApJS..222....8D} for single stars with $\rm [Fe/H]=-0.25$ at 1 Gyr, 2 Gyr, 3 Gyr, 5 Gyr and 10 Gyr are shown for comparison. The average uncertainties are shown by the black error bars. }
    \label{fig:fig10}
\end{figure*}

Figure \ref{fig:fig10} shows the distribution of the spectroscopically classified systems in $\rm \log (g)$ and $\rm [Fe/H]$ versus temperature. For comparison, we show MESA Isochrones and Stellar Tracks (MIST) isochrones \citep{2016ApJ...823..102C,2016ApJS..222....8D} for single stars with $\rm [Fe/H]=-0.25$ at 1 Gyr, 2 Gyr, 3 Gyr, 5 Gyr and 10 Gyr. The metallicity was chosen to match the median of the early-type systems. While using single star isochrones to interpret binaries combined with spectroscopic data models designed for single stars is risky, it is worth remembering that these systems basically have a single temperature. The most interesting feature of Figure \ref{fig:fig10} is probably that the systems have significantly lower $\rm \log (g)$ than expected for main-sequence stars. For the higher mass systems, this could be due to evolution, but the effect is present even for the lower mass systems which should not have had time to evolve. The offset seems to be largest near the break temperature and smallest at higher and lower temperatures. \citet{2018MNRAS.476..528E} did find that $\rm \log (g)$ was moderately underestimated in fits of single star models to semi-empirical binary models and we also find that detached binaries have similar offsets in $\rm \log (g)$. Non-variable single stars in the LAMOST survey have $\rm \log (g)$ values consistent with models of main-sequence stars, so either the $\rm \log (g)$ values are more biased than expected from \citet{2018MNRAS.476..528E} or there is a genuine difference. There also appears to be a weak trend of the systems having higher metallicities at lower temperatures.

\begin{figure}
	\includegraphics[width=0.5\textwidth]{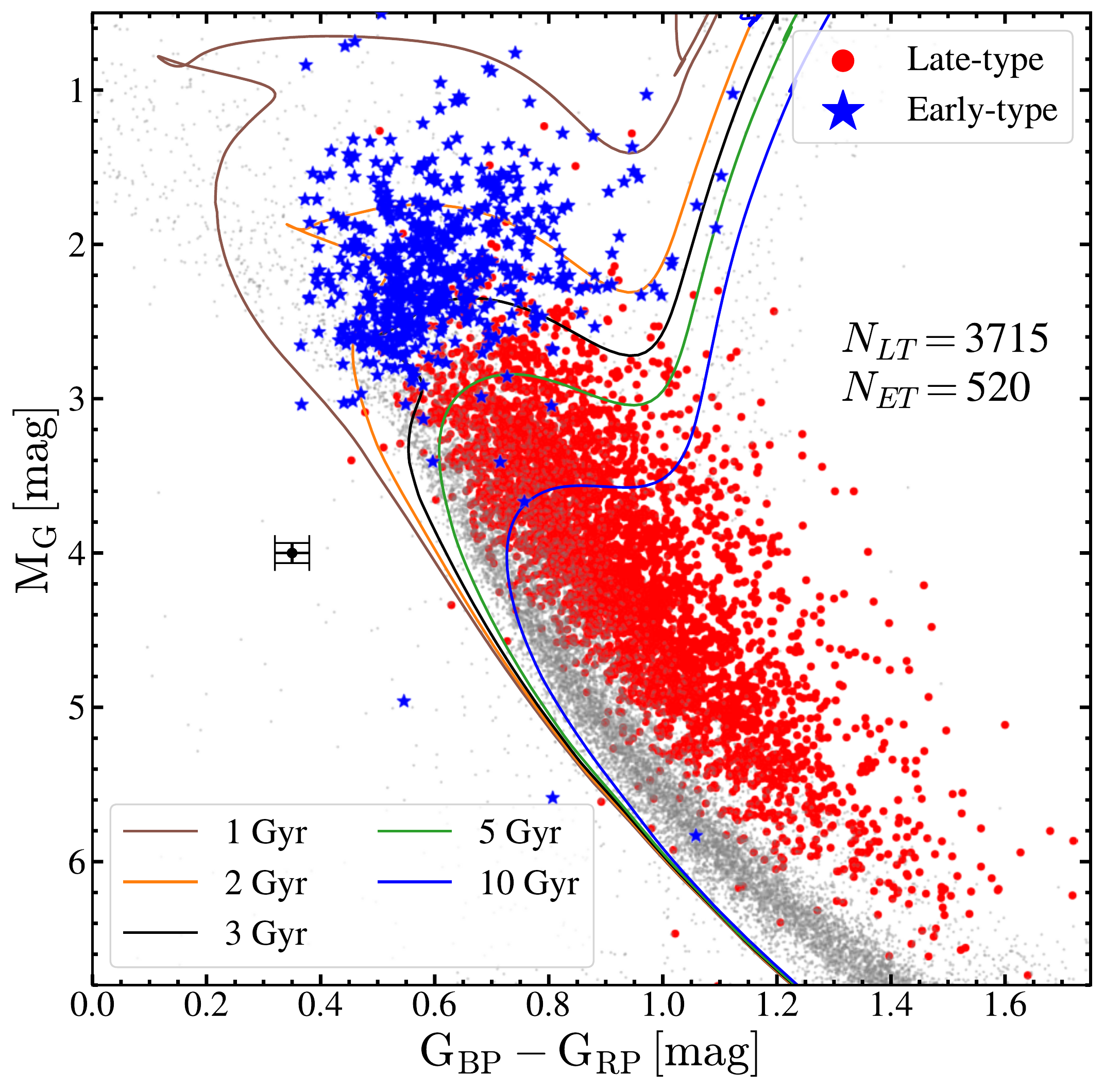}
    \caption{Gaia DR2 color-magnitude diagram for a sample of the early-type (blue) and late-type (red) EW binaries with $\rm Prob>0.90$, $A_V<1$ mag and parallaxes better than $10\%$, which were separated in period-temperature space using Equation \ref{eq:tefflogp}. A sample of nearby sources with good parallaxes and photometry is shown in gray. MIST isochrones \citep{2016ApJ...823..102C,2016ApJS..222....8D} for single stars with $\rm [Fe/H]=-0.25$ at 1 Gyr, 2 Gyr, 3 Gyr, 5 Gyr and 10 Gyr are shown for comparison. The average uncertainties are shown by the black error bars. Solar metallicity MIST isochrones lie on the gray points.}
    \label{fig:fig11}
\end{figure}

Figure \ref{fig:fig11} shows the early-type and late-type contact binaries with $A_V<1$ mag and parallaxes better than $10\%$ in a Gaia DR2 color-magnitude diagram (CMD) after correcting for interstellar extinction.  A sample of nearby sources with good parallaxes and photometry is shown in the background. The isochrones do not track the main-sequence of the nearby stars as these have near-Solar metallicities while the isochrones are for a lower metallicity ($\rm [Fe/H]=-0.25$). As expected, both groups of contact binaries are more luminous than stars on the main sequence and the early-type systems are more luminous than the late-type systems due to their higher masses.

\begin{figure*}
	\includegraphics[width=\textwidth]{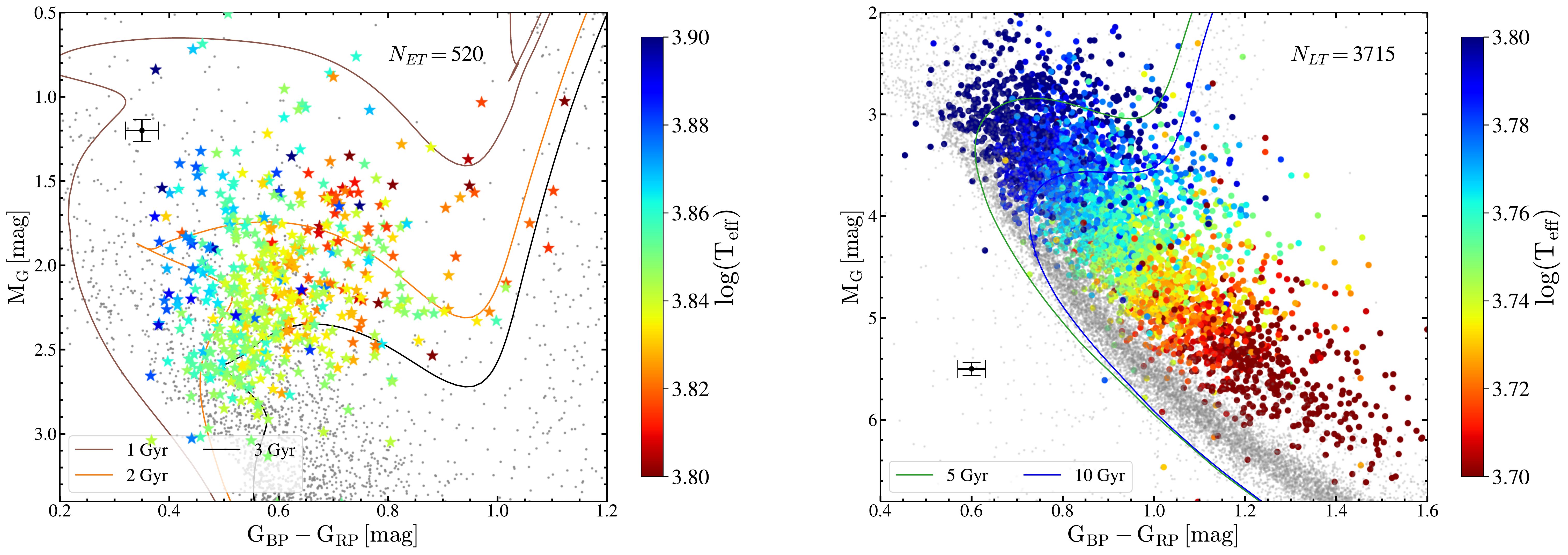}
    \caption{Gaia DR2 color-magnitude diagrams for the early-type (left) and late-type (right) EW binaries shown in Figure \ref{fig:fig11}, colored by $\log(\rm T_{eff})$. The colorbars show the same dynamic range in $\log(\rm T_{eff})$ of 0.1 dex. A sample of nearby sources with good parallaxes and photometry is shown in gray. MIST isochrones \citep{2016ApJ...823..102C,2016ApJS..222....8D} for single stars with $\rm [Fe/H]=-0.25$ at 1 Gyr, 2 Gyr, 3 Gyr, 5 Gyr and 10 Gyr are shown for comparison. The average uncertainties are shown by the black error bars. }
    \label{fig:fig12}
\end{figure*}

Figure \ref{fig:fig12} shows the same early-type and late-type sources in Figure \ref{fig:fig11}, colored by $\log(\rm T_{eff})$. The average temperature of the early-type binaries drop with the perpendicular distance from the main sequence, thus, early-type binaries are cooler if they are further from the main-sequence. The late-type binaries do not show a similar gradient in temperature with the perpendicular distance from the main-sequence. The temperature simply increases with luminosity. 

We use the empirical orbital period-mass relations derived in \citet{2008MNRAS.390.1577G} to derive estimates of the primary and secondary masses for the contact binaries with spectroscopic data. The masses derived from these relations have uncertainties of ${\sim}15\%$. The $1\%$-$99 \%$ quantile range of primary masses of the early-type and late-type systems were $1.27-3.29 M_{\odot}$ and $0.88-1.67 M_{\odot}$ respectively, whereas the distributions of secondary masses were $0.39-0.61 M_{\odot}$ and $0.33-0.44 M_{\odot}$ respectively. The total masses of the early-type and late-type systems were $1.66-3.90 M_{\odot}$ and $1.20-2.12 M_{\odot}$ respectively. The total masses of the W UMa binaries grow with the orbital period. These mass estimates are comparable to the theoretical estimates from \citet{2014MNRAS.437..185Y}. The median masses of the primaries for the early-type and late-type systems are $1.8 M_{\odot}$ and $1.2 M_{\odot}$ respectively. The masses of the primaries in the early-type (late-type) systems are consistent with their being above (below) the Kraft break. Simple main sequence lifetime arguments for the primaries of the early-type stars suggest that most of the sample should evolve off the main sequence at ${\sim}2$ Gyr, whereas the majority of the primaries of the late-type stars should evolve off the main sequence at ${\sim}6$ Gyr. This agrees with Figures \ref{fig:fig11} and \ref{fig:fig12}, where the early-type systems seem to be more evolved.

\section{Period-luminosity relations}
PLRs exist only for late-type EW systems and are based on small samples \citep{1994PASP..106..462R,2018ApJ...859..140C}. Here we derive PLRs using much larger samples and for both early and late-type systems.
We derive period-luminosity relations (PLRs) of the form \begin{equation}
M_\lambda=\rm A\log_{10}(P/0.5\,d)+B,
    \label{plrfit}
\end{equation} following the procedure in \citet{2019arXiv191014187J}. We corrected for interstellar extinction with the SFD estimate. We include systems with \verb"parallax"/\verb"parallax_error" $>20$, $\rm Prob>0.98$ and $A_V<1$ mag to reduce the uncertainty in the absolute magnitudes. We made an initial fit to each band, after which we removed outliers from the PLR fit by calculating the distance from the initial fit $$r=\sqrt{(\Delta \log_{10} \rm P)^2+(\Delta M_\lambda)^2},$$ where $$\Delta\log_{10}(\rm P)=\log_{10}( P_{fit}/P_{obs})$$ and$$\Delta M_\lambda=M_{\lambda,fit}-M_{\lambda,obs}.$$ Sources that deviated from this fit by $>2\sigma_r$ were removed. After removing these outliers, the parameters from the trial fit were then used to initialize a Monte Carlo Markov Chain sampler (MCMC) with 200 walkers, that were run for 20000 iterations using the MCMC implementation \verb"emcee" \citep{2013PASP..125..306F}. The errors in the PLR parameters were derived from the MCMC chains. Since we are using photometry obtained at random phases, these PLRs essentially correspond to the PLRs for the mean magnitudes of the binaries. However, they will have an additional scatter of $\sim 0.05$~mag because they are not individually averaged (or peak) magnitudes for each binary \citep{2018ApJ...859..140C}. For the early-type binaries, we only use the spectroscopic sample to avoid contamination by both late-type binaries and RRc variables, while for the late-type binaries, we augment the sample by including sources with $\rm \log (\rm P/d)<-0.4$ where there is no confusion between late-type and early-type contact binaries (Figure \ref{fig:fig7}). 

\begin{figure}
	\includegraphics[width=0.5\textwidth]{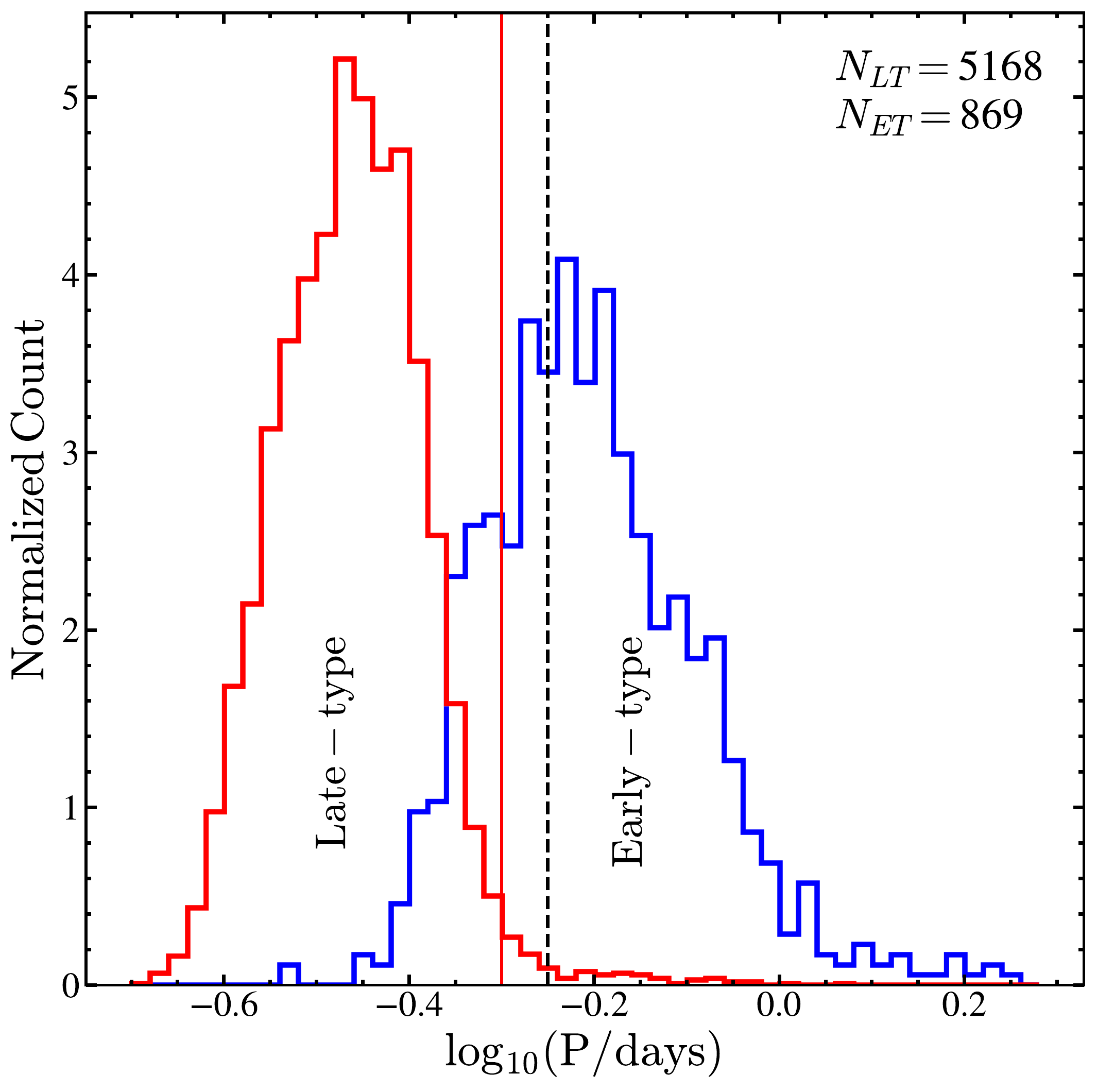}
    \caption{The distribution of orbital periods for the early-type (blue) and late-type (red) EW binaries which were separated in period-temperature space using Equation \ref{eq:tefflogp}. The usual period cut separating early-type and late-type systems of $\rm \log (\rm P/d) =-0.25$ is shown as a dashed black line. The suggested period cut of $\rm \log (\rm P/d) =-0.30$ for separating the systems is shown as a solid red line.}
    \label{fig:fig13}
\end{figure}

Given that most EW binaries do not have spectroscopic data to separate early-type and late-type systems based on equation \ref{eq:tefflogp}, we also derive PLRs for early-type and late-type systems separated by period. Figure \ref{fig:fig13} shows the period distribution of the early-type and late-type systems with spectroscopic information. To separate these systems based only on their orbital period, we use the cutoff of $\rm \log (\rm P/d) =-0.30$ at the minimum of the period distribution in Figure \ref{fig:fig5}. We chose the cut based on the orbital period distribution of the ${\sim}71,200$ EW binaries in ASAS-SN, instead of deriving a different cut from Figure \ref{fig:fig13} because it could be biased by selection effects in the spectroscopic sample. The usual cut of $\rm \log (\rm P/d) =-0.25$ excludes a significant number of shorter period early-type systems, but it does provide a clean sample of longer period early-type systems. The best-fit parameters, their uncertainties, the dispersion, and the number of sources used in the fit are listed in Table \ref{tab:ltfits} (late-type) and Table \ref{tab:etfits} (early-type) for the EW binaries separated in period-temperature space, and in Table \ref{tab:ltfitsp} (late-type) and Table \ref{tab:etfitsp} (early-type) for the EW binaries separated by period. 

\begin{figure}
	\includegraphics[width=0.5\textwidth]{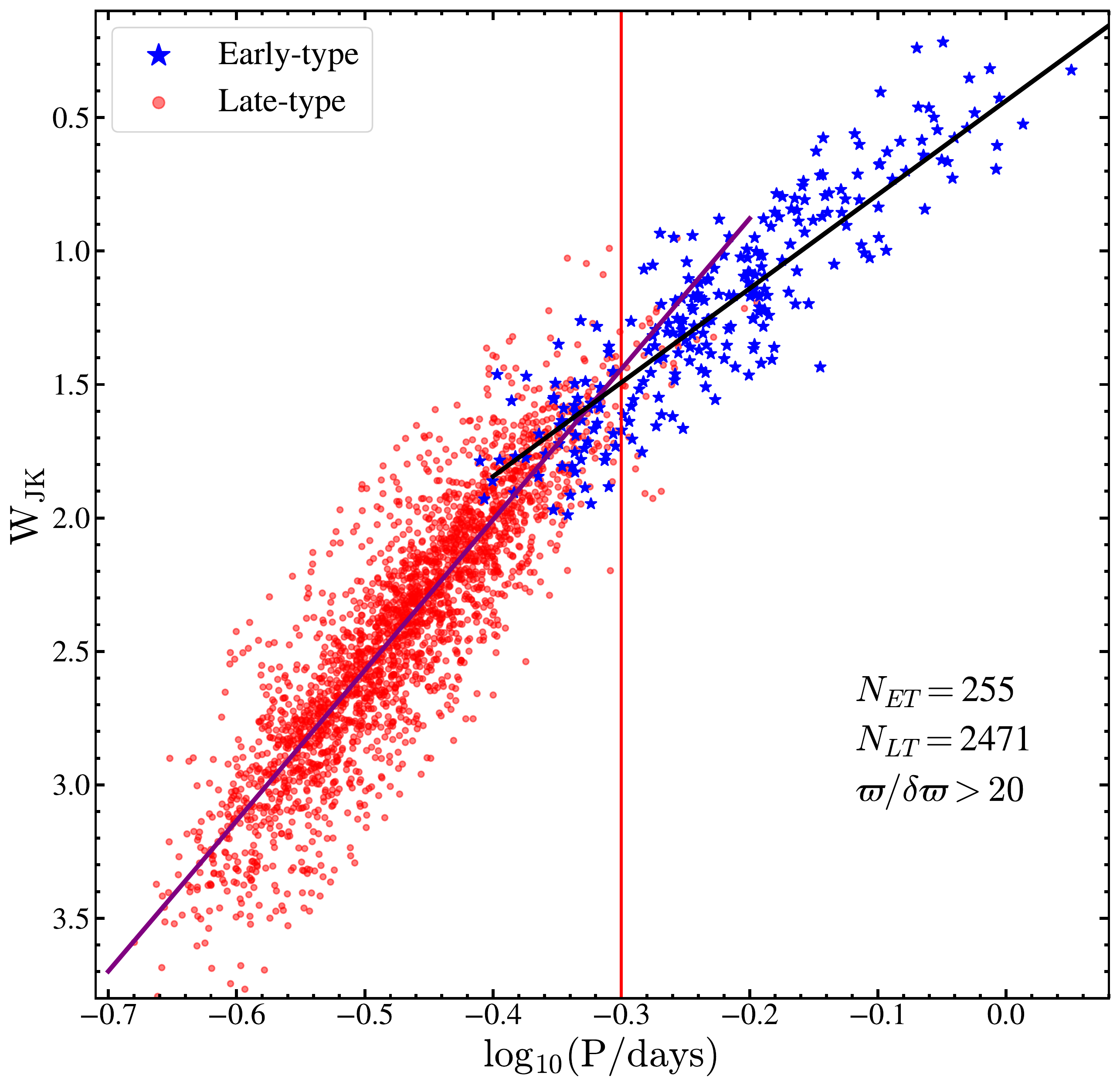}
    \caption{The Wesenheit $W_{JK}$ PLR diagram for the EW stars separated in period-temperature space using Equation \ref{eq:tefflogp}, with $\rm Prob>0.98$, $A_V<1$ mag and parallaxes better than $5\%$ after clipping outliers with dispersions $>3\sigma_r$ from the respective PLR fit. The fitted PLRs for the late-type and early-type contact binaries are shown as purple and black lines respectively. The suggested period cut of $\rm \log (\rm P/d) =-0.30$ for separating the systems is shown as a solid red line.}
    \label{fig:fig14}
\end{figure}

Figure \ref{fig:fig14} illustrates the PLRs of the late-type and early-type systems separated in period-temperature space. The PLR slopes are significantly different for the early-type and late-type binaries. The PLR slopes for the late-type systems separated in period-temperature space also differ by ${\lesssim}1\%$ from the PLR slopes of the late-type systems separated by period. The slopes for the late-type systems derived with these two samples are consistent given the errors. On the other hand, early-type systems have slopes that differ by ${\sim}1\%$ in the V-band and ${\sim}25\%$ in the $K_s$-band. The NIR PLR slopes for the early-type systems between the two ways of separating the EW sub-types are inconsistent given their uncertainties. The slopes of the PLR fits for the late-type binaries differ by $>3\sigma$ with those obtained by \citet{2018ApJ...859..140C}, but the disagreement is smaller for the NIR PLRs. For example, their V-band ($-9.14\pm0.40$) and $K_s$-band ($-5.95\pm0.21$) slopes are shallower than the slopes derived here by ${\sim}14\%$ and ${\sim}7\%$ respectively. In addition, at the median period $\rm P{\sim}0.34$ d of the late-type EW binaries, the PLR fits differ by ${\sim}0.3$ mag in the V-band and ${\sim}0.1$ mag in the $K_s$-band from those obtained by \citet{2018ApJ...859..140C}. These differences are smaller than the scatter in the PLRs at these bands. \citet{2018ApJ...859..140C} derived PLRs for the late-type contact binaries based on a small sample of only 183 nearby ($d<330$ pc) sources with an average extinction of $A_V=0.075\pm0.025\,\rm mag$, whereas our PLR fits are made to a much bigger sample (${\sim}61\times$) across a wider range in distance and $A_V$. There are no significant changes in the PLR fits if we restrict our sample to smaller extinctions ($A_V<0.5$ mag). Contact binaries are more diverse and span a much larger range in luminosity and temperature than classical pulsators like RR Lyrae and Cepheids, leading to more dispersion about the PLR.

\begin{table}
	\centering
	\caption{PLR parameters for the late-type contact binaries separated in period-temperature space}
	\label{tab:ltfits}
\begin{tabular}{rcccc}
		\hline
		 Band & A & B & $\sigma$ & N\\
		  & mag & mag & mag & \\		 
		\hline
		$W_{JK}$ & $-5.527 \pm 0.052$&  $1.469 \pm 0.049$ &  0.197& 11384\\
		$V$ & $-10.628 \pm 0.052$&  $2.360 \pm 0.054$ &  0.383 & 11134\\
		$G$ & $-10.173 \pm 0.047$&  $2.379 \pm 0.050$ &  0.337 & 11329\\
		$J$ & $-7.651 \pm 0.049$&  $1.763 \pm 0.054$ &  0.254 & 11363\\	
		$H$ & $-6.589 \pm 0.047$&  $1.623 \pm 0.048$ &  0.222 & 11414\\
		$K_s$ & $-6.381 \pm 0.051$&  $1.601 \pm 0.045$ &  0.214 & 11387\\	
		$W_1$ & $-6.369 \pm 0.051$&  $1.540 \pm 0.051$ &  0.199 & 11579\\		
\hline
\end{tabular}
\end{table}

\begin{table}
	\centering
	\caption{PLR parameters for the early-type contact binaries separated in period-temperature space}
	\label{tab:etfits}
\begin{tabular}{lcccc}
		\hline
		 Band & A & B & $\sigma$ & N\\
		  & mag & mag & mag & \\		 
		\hline
		$W_{JK}$ & $-3.616 \pm 0.048$&  $1.499 \pm 0.049$ &  0.140& 226\\
		$V$ & $-2.770 \pm 0.049$&  $2.407 \pm 0.051$ &  0.166 & 221\\
		$G$ & $-2.977 \pm 0.047$&  $2.426 \pm 0.049$ &  0.154 & 228\\
		$J$ & $-3.512 \pm 0.050$&  $1.799 \pm 0.049$ &  0.144 & 226\\	
		$H$ & $-3.535 \pm 0.050$&  $1.650 \pm 0.050$ &  0.145 & 231\\
		$K_s$ & $-3.520 \pm 0.051$&  $1.612 \pm 0.043$ &  0.142 & 228\\	
		$W_1$ & $-3.565 \pm 0.056$&  $1.583 \pm 0.045$ &  0.129 & 234\\		
\hline
\end{tabular}
\end{table}

\begin{table}
	\centering
	\caption{PLR parameters for the late-type contact binaries separated by period}
	\label{tab:ltfitsp}
\begin{tabular}{lcccc}
		\hline
		 Band & A & B & $\sigma$ & N\\
		  & mag & mag & mag & \\		 
		\hline
		$W_{JK}$ & $-5.523 \pm 0.049$&  $1.482 \pm 0.059$ &  0.195& 11118\\
		$V$ & $-10.531 \pm 0.047$&  $2.373 \pm 0.047$ &  0.378 & 10843\\
		$G$ & $-10.076 \pm 0.047$&  $2.410 \pm 0.057$ &  0.331 & 11007\\
		$J$ & $-7.678 \pm 0.046$&  $1.753 \pm 0.051$ &  0.251 & 11089\\	
		$H$ & $-6.612 \pm 0.048$&  $1.619 \pm 0.054$ &  0.220 & 11148\\
		$K_s$ & $-6.402 \pm 0.044$&  $1.581 \pm 0.053$ &  0.212 & 11123\\	
		$W_1$ & $-6.359 \pm 0.055$&  $1.545 \pm 0.055$ &  0.197 & 11300\\		
\hline
\end{tabular}
\end{table}

\begin{table}
	\centering
	\caption{PLR parameters for the early-type contact binaries separated by period}
	\label{tab:etfitsp}
\begin{tabular}{lcccc}
		\hline
		 Band & A & B & $\sigma$ & N\\
		  & mag & mag & mag & \\		 
		\hline
		$W_{JK}$ & $-2.640 \pm 0.051$&  $1.403 \pm 0.059$ &  0.258& 1199\\
		$V$ & $-2.753 \pm 0.049$&  $2.328 \pm 0.047$ &  0.351 & 1176\\
		$G$ & $-2.772 \pm 0.049$&  $2.379 \pm 0.051$ &  0.334 & 1186\\
		$J$ & $-2.635 \pm 0.051$&  $1.716 \pm 0.050$ &  0.284 & 1189\\	
		$H$ & $-2.675 \pm 0.048$&  $1.575 \pm 0.053$ &  0.271 & 1198\\
		$K_s$ & $-2.656 \pm 0.051$&  $1.528 \pm 0.054$ &  0.269 & 1203\\	
		$W_1$ & $-2.709 \pm 0.055$&  $1.487\pm 0.060$ &  0.250 & 1192\\		
\hline
\end{tabular}
\end{table}

\section{Conclusions}

We analyzed a sample of 71242 W UMa eclipsing binaries, including 12584
new discoveries, in the ASAS-SN V-band catalog of variable stars, taking
advantage of their Gaia DR2 parallaxes \citep{2018arXiv180409365G} and the spectroscopic
temperatures, metallicities and $\log(\rm g)$ estimates from primarily LAMOST,
but also GALAH, RAVE and APOGEE, for 7169 of the stars. The large sample size and
(in particular) the spectroscopic temperatures lead to a much clearer
view of the dichotomy between early and late-type systems.  The period
distribution has a clear minimum at $\rm \log (\rm P/d) =-0.30$, making it a better
separator between the two populations than the standard of $\rm \log (\rm P/d) =-0.25$.
The period-luminosity relation also has a distinct break in its slope at the
same period.  

The distinction between the populations is even clearer in the space of period and effective temperature, where there is essentially a gap along the line $\rm T_{eff}=6710K-1760K\,\log(P/0.5\,d)$.  There are no strong distinctions between the populations in metallicity or log(g), nor strong correlations within the populations with these properties. Early-type systems are hotter at shorter orbital periods and get cooler as the orbital period increases. The total mass of the EW binaries increases with their orbital period. Thus more massive early-type EW are cooler and likely more evolved than the less massive early-type EW that are hotter. 

With the larger number of systems and a clearer separation of the two classes, we then derive revised period-luminosity relations for the late-type and early-type EW binaries in the $W_{JK}$, $V$, Gaia DR2 $G$, $J$, $H$, $K_s$ and $W_1$ bands for contact binaries both separated spectroscopically and by period. The slopes of the late-type PLRs for the two ways of dividing the systems were consistent with each other given the uncertainties, but the slopes of the early-type PLRs differ by ${\sim}25\%$ in the NIR. The slopes we find for late-type PLRs differ significantly (${\sim}10\%$) from the existing PLRs for late-type EW binaries given their uncertainties. This is likely due to the far smaller samples used by previous studies to derive PLRs.

The Kraft break appears to determine the observed dichotomy of the contact binaries. Stars lose angular momentum inefficiently above the Kraft break, making it unlikely that angular momentum loss is sufficient to bring the early-type systems into contact. Thus, early-type systems form due to stellar evolution and the subsequent expansion of a more massive component that is above the Kraft break (${\sim}1.3 M_{\odot}$). For the late-type systems, the primary is below the Kraft break, and the late-type systems can come into contact due to efficient angular momentum loss during the detached phase. The positions of the EW binaries on a Gaia DR2 color-magnitude diagram are consistent with early-type EW binaries being younger and more evolved than the late-type systems. Late-type EW binaries appear to be main sequence binaries and the vast majority of these appear older than 5 Gyr. This is consistent with standard models for the formation and evolution of these systems \citep{2014MNRAS.437..185Y,2014MNRAS.438..859J}.

\section*{Acknowledgements}

We thank the referee, Dr. Diana Kjurkchieva, for the very useful comments that improved our presentation of this work. We thank Dr. Jennifer Johnson for useful discussions on this manuscript. We thank the Las Cumbres Observatory and its staff for its continuing support 
of the ASAS-SN project. We also thank the Ohio State University College of Arts 
and Sciences Technology Services for helping us set up and maintain the ASAS-SN 
variable stars and photometry databases.

ASAS-SN is supported by the Gordon and Betty Moore
Foundation through grant GBMF5490 to the Ohio State
University, and NSF grants AST-1515927 and AST-1908570. Development of
ASAS-SN has been supported by NSF grant AST-0908816,
the Mt. Cuba Astronomical Foundation, the Center for Cosmology 
and AstroParticle Physics at the Ohio State University, 
the Chinese Academy of Sciences South America Center
for Astronomy (CAS- SACA), the Villum Foundation, and
George Skestos. 

KZS and CSK are supported by NSF grants AST-1515927, AST-1814440, and 
AST-1908570. BJS is supported by NSF grants AST-1908952, AST-1920392, 
and AST-1911074. TAT acknowledges support from a Simons Foundation 
Fellowship and from an IBM Einstein Fellowship from the Institute for 
Advanced Study, Princeton. Support for JLP is provided in part by the
Ministry of Economy, Development, and Tourism's Millennium Science 
Initiative through grant IC120009, awarded to The Millennium Institute 
of Astrophysics, MAS. Support for MP and OP has been provided by 
INTER-EXCELLENCE grant LTAUSA18093 from the Czech Ministry of 
Education, Youth, and Sports. The research of OP has also been 
supported by Horizon 2020 ERC Starting Grant ``Cat-In-hAT'' 
(grant agreement \#803158) and PRIMUS/SCI/17 award from Charles 
University. This work was partly supported by NSFC 11721303.

This work has made use of data from the European Space Agency (ESA)
mission {\it Gaia} (\url{https://www.cosmos.esa.int/gaia}), processed by
the {\it Gaia} Data Processing and Analysis Consortium. This publication makes 
use of data products from the Two Micron All Sky Survey, as well as
data products from the Wide-field Infrared Survey Explorer.
This research was also made possible through the use of the AAVSO Photometric 
All-Sky Survey (APASS), funded by the Robert Martin Ayers Sciences Fund. 

This research has made use of the VizieR catalogue access tool, CDS, Strasbourg, France. 
This research also made use of Astropy, a community-developed core Python package for 
Astronomy (Astropy Collaboration, 2013).












\bsp	
\label{lastpage}
\end{document}